\documentclass[sigconf]{acmart}

\AtBeginDocument{%
  }

\usepackage{multirow}
\usepackage{url}
\usepackage{float}
\usepackage[skip=2pt]{caption}
\usepackage{enumitem}

\usepackage{xcolor}
\definecolor{darkblue}{rgb}{0.733,0.870,0.984} 
\definecolor{lightblue}{rgb}{0.392,0.710,0.964}   

\definecolor{bgcolor}{RGB}{230, 240, 255}

\begin{document}

\title{Practical Feasibility of Sustainable Software Engineering Tools and Techniques}

\author{Satwik Ghanta}
\affiliation{%
  \institution{University of Glasgow}
  \city{Glasgow}
  \country{United Kingdom}}
\email{y.ghanta.1@research.gla.ac.uk}

\author{Peggy Gregory}
\affiliation{%
  \institution{University of Glasgow}
  \city{Glasgow}
  \country{United Kingdom}}
\email{Peggy.Gregory@glasgow.ac.uk}

\author{G\"{u}l \c{C}al{\i}kl{\i}}
\affiliation{%
  \institution{University of Glasgow}
  \city{Glasgow}
  \country{United Kingdom}}
\email{HandanGul.Calikli@glasgow.ac.uk}

\renewcommand{\shortauthors}{Ghanta et al.}

\begin{abstract}
While Sustainable Software Engineering (SSE) tools are widely studied in academia, their practical feasibility in industrial workflows, particularly in regulated environments, remains poorly understood.
This study investigates how software practitioners perceive the feasibility of existing SSE tools and techniques, and examines the technical, organizational, and cultural factors shaping their adoption in practice.
We identified prominent categories of SSE tools targeting energy consumption, green refactoring, and workload management, and evaluated them along three practitioner-relevant dimensions: installation, input requirements, and output formats. These were presented through an interactive web application and explored in workshops with 16 practitioners from a regulated financial-sector organization, followed by a survey of 27 software practitioners.
Our findings suggest that the practitioners strongly favored tools that integrate into existing IDEs or pipelines, require minimal and locally scoped data access, and provide interpretable, actionable outputs such as dashboards or automated refactoring suggestions. In regulated settings, compliance requirements, approval processes, and time constraints significantly shaped feasibility perceptions.
Our contribution lies in providing empirical evidence of these preferences alongside other factors that affect regulated industrial contexts. The findings offer actionable guidance for designing SSE tools that better align with real-world development workflows and organizational constraints.
\end{abstract}

\ccsdesc[500]{Social and professional topics~Sustainability}
\ccsdesc[500]{Software and its engineering~Software maintenance tools}
\ccsdesc[500]{Social and professional topics~Project and people management}
\ccsdesc[500]{Social and professional topics~Socio-technical systems}
\ccsdesc[100]{Social and professional topics~Computing education}

\keywords{Sustainability, Sustainability in software engineering, Sustainability tools, Socio-technical systems, Software engineering teams, Green software, Organizational sustainability, Green organization}

\maketitle

\section{Introduction}

Sustainable Software Engineering (SSE) has gained growing attention in recent years as the software industry faces increasing pressure to address various dimensions of sustainability, driven by factors such as competitive positioning, regulatory demands, and organizational reputation~\cite{ghanta_perspectives_2025}. While the literature proposes a wide range of tools and techniques aimed at improving energy efficiency, promoting responsible coding practices, and supporting developer well-being, etc., their adoption in real-world settings remains limited~\cite{ghanta_perspectives_2025,chitchyan_sustainability_2016,groher_interview_2017}. Despite academic interest, many of these approaches are developed and tested in controlled or idealized environments, and rarely transition into everyday software development workflows. As a result, there is a growing disconnect between what is proposed in research and what is feasible or acceptable in practice.

Another contributing factor to this gap is that software practitioners often lack knowledge about which SSE tools exist, how they work, and how they might fit within their development environments~\cite{ghanta_perspectives_2025, karita_towards_2022, groher_interview_2017, chitchyan_sustainability_2016, manotas_empirical_2016, souza_defining_2014}. Even when awareness exists, sustainability is frequently de-prioritized due to misaligned incentive structures that emphasize short-term delivery over long-term impact~\cite{ghanta_perspectives_2025}. As a result, the enablers and barriers that influence the practical uptake of SSE tools and techniques remain poorly understood. Addressing this gap requires clearly outlining the technical, organizational, and cultural factors that influence the feasibility of adopting SSE tools in practice, where organizational factors are the formal processes and structures of the company, while cultural factors are the shared values and everyday practices of its people.

To address these challenges, we explore how software practitioners perceive the applicability of existing SSE tools and the factors shaping their views. Our study focuses on three key areas where sustainability intersects with software development workflows: energy consumption, green refactoring, and workload management. Rather than introducing new tools, we examine how existing ones are perceived in terms of usefulness, integration effort, and alignment with development environments. By grounding our investigation in the lived experiences of professionals in software-intensive organizations, we contribute practical insights into the conditions under which SSE tools can be meaningfully adopted.

\subsection{Data Availability}
The replication package~\cite{anonymous_2025_18337075} includes literature review search strings; pre- and post-screening lists of matching literature; workshop and survey artifacts (invitation sheet, participant demographics, questionnaires and responses); and the web application's codebase with setup documentation. It also contains Python notebooks for statistical analyses of workshop and survey responses, and the codebook for workshop data (qualitative) analyses. Due to organizational restrictions, the organization was anonymized and audio transcripts were not shared.

\begin{figure*}
    \includegraphics[width=0.7\textwidth]{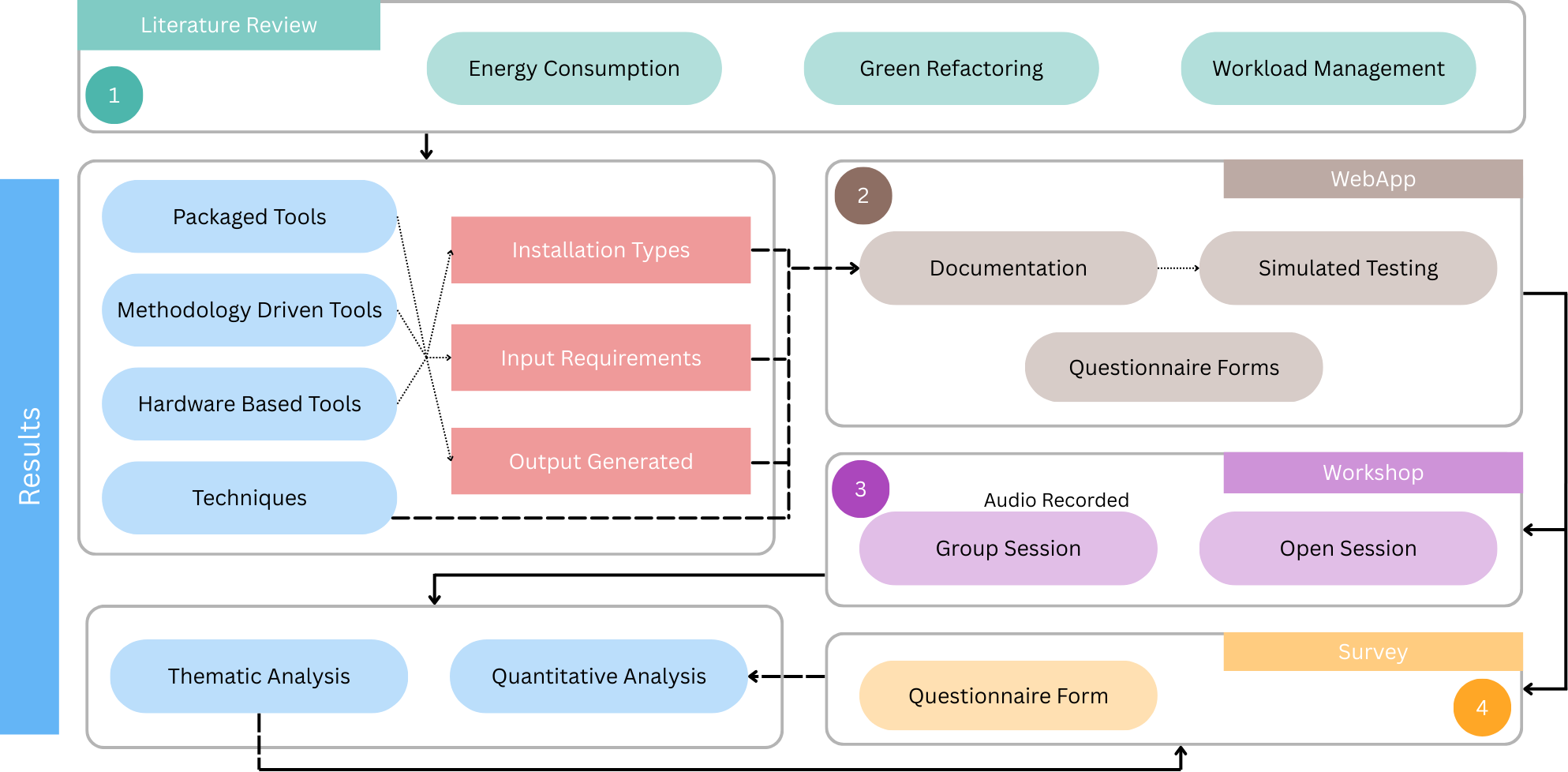}
    \caption{Mixed-methods research design and flow, starting with  \textcircled{1} a literature review to identify SSE tool categories (energy consumption, green refactoring, and workload management). \textcircled{2} A custom web application was used during  \textcircled{3} the workshop and \textcircled{4} the survey. 
    We conducted thematic and statistical analyses on workshop and survey data, respectively.
    }
    \label{fig:flow}
\end{figure*}

\section{Related Work}

A recurring theme in the literature is the difficulty of embedding sustainability practices into organizational contexts. Bamiduro et al.~\cite{bamiduro_challenges_2025} identified key barriers through interviews with practitioners, such as lack of prioritization, inadequate measurement tools, and organizational culture. Similarly, Ournani et al.~\cite{ournani_reducing_2020} interviewed experienced developers and found that, while energy consumption was acknowledged, concrete practices were poorly understood and rarely acted upon. Montes et al.~\cite{montes_factors_2025} investigated factors influencing software engineers' well-being and showed that cultural norms and organizational support strongly shape sustainable work practices. All works highlight organizational and cultural obstacles to adoption, but they do not directly evaluate the usability or integration of specific tools. Our study builds on these insights by examining not only organizational challenges but also how technical and cultural factors shape practitioner evaluations of concrete tool categories.

Another line of research has focused on awareness-raising interventions. Jagroep et al.~\cite{jagroep_awakening_2017} conducted a multiple-case study in which an energy dashboard was deployed to track energy consumption across sprints. While the dashboard triggered discussion and increased awareness, participants were divided on its long-term practicality. Rani et al.~\cite{rani_energy_2024} extended this line of inquiry by porting energy patterns from the mobile to the web domain, combining interviews with developers and empirical testing. They showed that many patterns were poorly understood and practical benefits were often uncertain. These works emphasize the role of awareness but do not investigate how practitioners perceive broader classes of SSE tools, a gap our study addresses by systematically surfacing multiple types of tools and eliciting structured feedback.

Complementing these empirical efforts, literature reviews and mapping studies have provided overviews of sustainability practices. Mourao et al.~\cite{mourao_green_2018} synthesized 75 studies through a systematic mapping study, showing growing academic interest but highlighting the lack of empirical validation and industry alignment. Junger et al.~\cite{junger_potentials_2024} focused on green coding, compiling practices and recommendations and considering how they might be integrated into industrial processes and education. Both reviews underscore the richness of proposed approaches but also the limited evidence of their feasibility in practice. Our study directly responds to this gap by exposing practitioners to existing tools and techniques and analyzing their adoption potential in real-world settings.

Finally, there are educational approaches aimed at sustainability through better coding practices. Luburic et al.~\cite{luburic_clean_2022} developed an intelligent tutoring system for teaching clean code and refactoring, demonstrating its potential for training but also showing the challenges of measuring impact. While their focus was on education rather than industry adoption, their findings resonate with broader concerns about the need for practical training and contextual fit. In contrast, our work does not design new educational tools but instead evaluates how existing SSE tools align with professional practices and organizational constraints.

In summary, previous research has revealed the barriers, awareness gaps, and conceptual richness of sustainable software engineering, but the practical feasibility of tools remains unexplored across dimensions such as installation, input requirements, output usefulness, and adoption factors. By combining a literature review with workshop and survey, our study contributes empirical evidence on practitioners' perceptions on the SSE tools' feasibility, providing actionable insights for their design and adoption.

\section{Methodology}
We employed an exploratory sequential mixed-methods design~\cite{storey_guiding_2025}. This section presents our research questions and provides details of each phase of our study design as depicted in Figure~\ref{fig:flow}. Ethical approval for this study was obtained prior to data collection from the University of Glasgow ethics committee.

\subsection{Research Questions}
\smallskip

\begin{description}[leftmargin=0.15cm]

  \item \textbf{RQ$_1$. \emph{How do software practitioners perceive the feasibility of different types of SSE tools?}}
 \\This question explores practitioners' interpretations, reactions, and experiences when engaging with various types of SSE tools across three technical dimensions (i.e., installation, input, and output formats), focusing on how these dimensions align with their practices, work environments, and professional expectations.

\end{description}

\begin{description}[leftmargin=0.15cm]

  \item \textbf{RQ$_2$. \emph{What factors influence perceptions of SSE tools' feasibility in software development workflows?}}
 \\This question explores organizational (formal processes and organization structure) and cultural (shared values and daily practices) enablers and barriers affecting SSE tools' adoption into workflows.

\end{description}

\subsection{Relevant Areas}
In this study, we focus on three main areas: \textit{energy consumption}, \textit{green refactoring}, and \textit{workload management}. We selected these areas based on their strong presence in existing research and the practical concerns raised by the participating organization.

Energy consumption and green refactoring are well-established focus areas within SSE literature~\cite{mourao_green_2018}. Numerous tools and techniques have been developed to monitor and optimize the energy efficiency of software systems, either through runtime analysis or by guiding developers toward more sustainable code practices. These areas offer a rich foundation for investigating real-world feasibility, given the availability of tools and prior empirical studies.

In contrast, developer workload management is comparatively underexplored in the context of sustainability, despite its relevance to individual and organizational well-being. Our inclusion of this topic is motivated by findings from a previous research study with in the financial industry~\cite{ghanta_perspectives_2025}, in which participants highlighted workload-related stress and inefficiencies as critical barriers to sustainable software development. In this previous study participants' feedback emphasized the need for tools and practices that not only support environmental goals but also address human-centric sustainability concerns.

\subsection{Literature Review}
To investigate perceptions of SSE tools and techniques, we began with a Focused Targeted Literature Review (FTLR). This approach, also sometimes described as a targeted or focused review~\cite{TLR_types_2023, smela_rapid_2023}, emphasizes finding the most relevant practice-oriented evidence rather than aiming for exhaustive coverage. Unlike systematic reviews, which aim for complete coverage, an FTLR narrows the scope to highly relevant publications and applies pragmatic screening criteria to generate actionable insights within a specific domain~\cite{healthcare_types_2022}. An FTLR was an essential step to identify the current landscape of SSE-related tools, their characteristics, and the extent to which prior work has evaluated their practical applicability. Rather than proposing new tools, our aim was to understand how existing solutions align with industry needs, particularly in terms of integration into development workflows, usability, and developer perception.

Prior reviews of SSE research have frequently noted a theoretical bias, with limited empirical validation of sustainability interventions~\cite{mcguire_sustainability_2023}. Moreover, much of the literature concentrates on ecological or technical sustainability, often neglecting the organizational and cultural realities that determine adoption.

\subsubsection{\textbf{Search Source}} We conducted an FTLR using the Scopus digital database~\cite{scopus}, selected for its comprehensive indexing of computer science publications and consistent query syntax support. Only articles published between 2015 and 2025 were considered. The aim was not to compile an exhaustive list of tools, but to surface recurring categories of approaches relevant to practitioners and to identify example tools in each category. We restricted the subject area to Computer Science (COMP) to maintain a focus on software-specific tools and practices.

\subsubsection{\textbf{Search Strategy}} To capture literature across key sustainability domains, we formulated three distinct query clusters, each aligned with a specific area of interest: Energy Consumption, Green Refactoring, and Developer Workload. Each query used combinations of \textit{TITLE-ABS-KEY} fields with proximity operators (e.g., W/3) and wildcards (e.g., *) to capture relevant variations in terminology. Example terms included \textit{``developer W/3 workload", ``energy efficien* W/3 tool*", and ``sustainab* W/3 cod* W/3 practice*''}. Full query examples are provided in the replication package~\cite{anonymous_2025_18337075}.

\subsubsection{\textbf{Inclusion Criteria}} To reduce noise and enhance specificity, we applied the following inclusion criteria across all areas:
\begin{itemize}[leftmargin=*]
    \item The paper must explicitly relate to software engineering or software development.
    \item The focus must include at least one sustainability dimension (Environmental, Economic, Social, Technical, Individual).
    \item The paper must describe or evaluate a tool, technique, or practice relevant to software professionals.
\end{itemize}

\subsubsection{\textbf{Screening}} Initial search results were filtered using Scopus' metadata fields. Duplicates were automatically removed, and titles and abstracts were screened manually for relevance. We excluded papers that:
\begin{itemize}[leftmargin=*]
    \item Focused on general sustainability but lacked actionable tool/technique proposals.
    \item Provided conceptual models without technical implementation and/or evaluation.
\end{itemize}
The purpose of screening was not to identify the ``best" tools, but to ensure the set was representative of the range of installation modes, input types, and output forms described in the literature. Table \ref{tab:litreview_search} reports the number of papers retained after the initial screening. The full list of screened literature is provided in the replication package\footnote{Replication package~\cite{anonymous_2025_18337075} contains \texttt{Search String} folder with pre- and post-screening files.}.

\begin{table}[h]
\centering
\caption{Literature search results across areas (Scopus).}
\begin{tabular}{l c c}
\hline
\textbf{Topic} & \textbf{Initial hits} & \textbf{Initial screening} \\
\hline
Developer workload       & 1488 & 93 \\
Green Refactoring        & 408  & 282 \\
Energy consumption       & 499  & 68 \\
\hline
\end{tabular}
\label{tab:litreview_search}
\end{table}

\begin{figure*}
    \centering
    \includegraphics[width=0.8\textwidth]{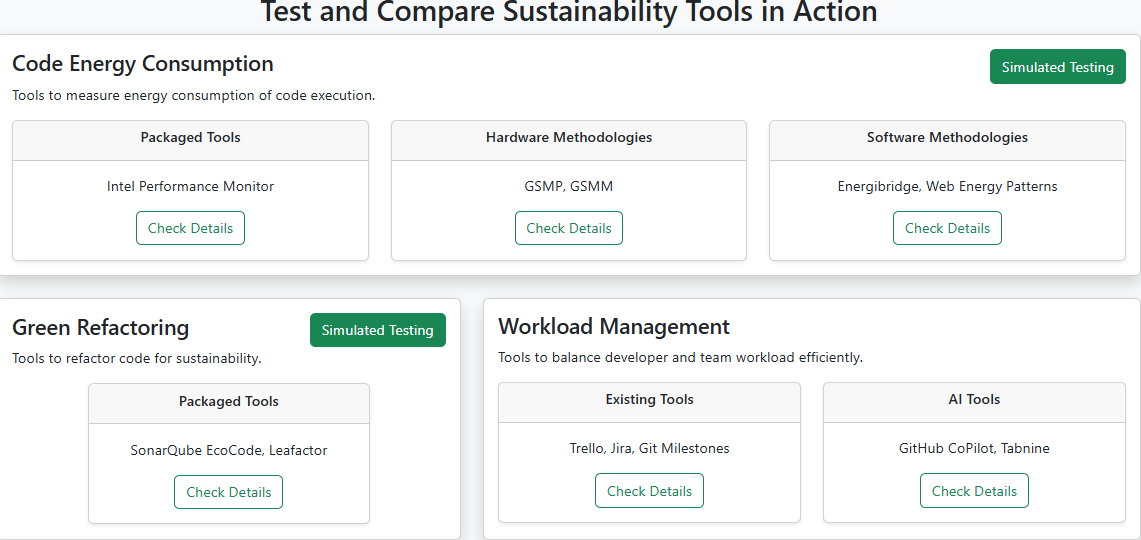}
    \caption{UI screen: Tools}
    \label{fig:ui_tools_all}
\end{figure*}

\begin{table*}[ht]
\centering
\resizebox{\textwidth}{!}{%
\begin{tabular}{lllll}
\hline
\textbf{Type} & \textbf{Example Tools} & \textbf{Installation Method} & \textbf{Input Data} & \textbf{Default Output Type} \\
\hline
\multicolumn{5}{l}{\textbf{Energy Consumption}} \\
Packaged Tools & Intel Performance Counter Monitor & Direct Installation & Code File & Raw Metrics \\
Hardware Methodologies & GSMP, GSMM & Hardware Device & Codebase access & Raw Metrics \\
Software Methodologies & Energibridge, Web Energy Patterns Study & Manual Building & Code File or Codebase access & Raw Metrics and Graphical \\  \hline
\multicolumn{5}{l}{\textbf{Green Refactoring}} \\
Packaged Tools & SonarQube EcoCode, Leafactor & IDE Plugin & Code File & Automated Changes \\ \hline
\multicolumn{5}{l}{\textbf{Workload Management}} \\
Existing Tools & Trello, Jira, Git Milestones & Direct Installation or Cloud & Tasks, commits, milestones & Graphical \\
AI Tools & GitHub CoPilot, Tabnine & IDE plugin or Cloud & Code File or developer specific input & Automatic Changes or Suggestions \\ \hline
\end{tabular}%
}
\caption{Overview of SSE Tools Presented}
\label{tab:sustainability_tools}
\end{table*}

\subsubsection{\textbf{Analysis}} For each included paper, we extracted: installation method (e.g., standalone, plugin, web-based), required input data (e.g., code, database), and output type (e.g., dashboard, numerical metrics). Rather than cataloging individual tools, this study abstracts the tools into categories based on how practitioners encounter and adopt them in practice. Through the focused literature review, we observed that tools consistently differ along installation dimension rather than underlying measurement mechanism. As a result, they were grouped into \textbf{(i)} packaged tools that are directly installable and ready to use, \textbf{(ii)} software methodologies that require building, scripting, or manual configuration from source, and \textbf{(iii)} hardware-based approaches that rely on physical devices or hardware counters.

\begin{figure*}
    \begin{tabular}{ll}
    \centering
    \includegraphics[width=0.45\textwidth]{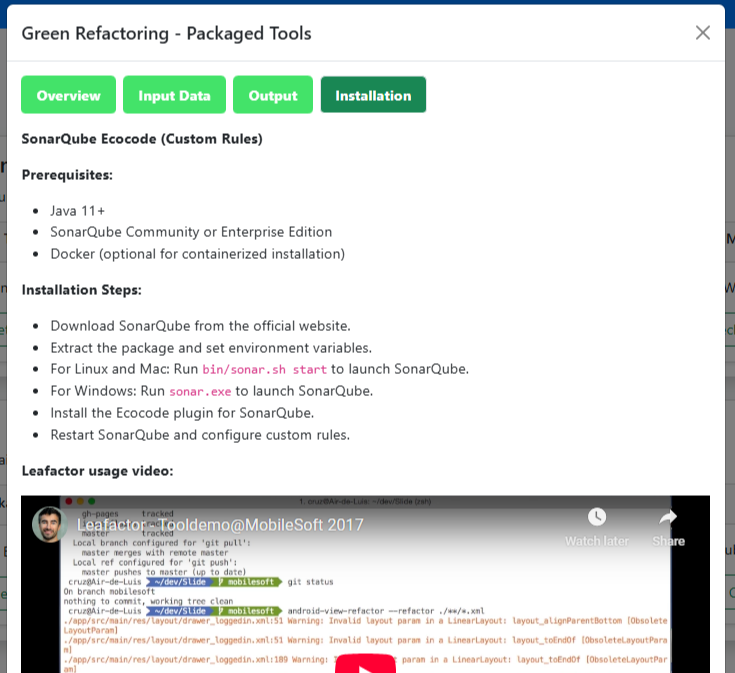}
    &
    \includegraphics[width=0.45\textwidth]{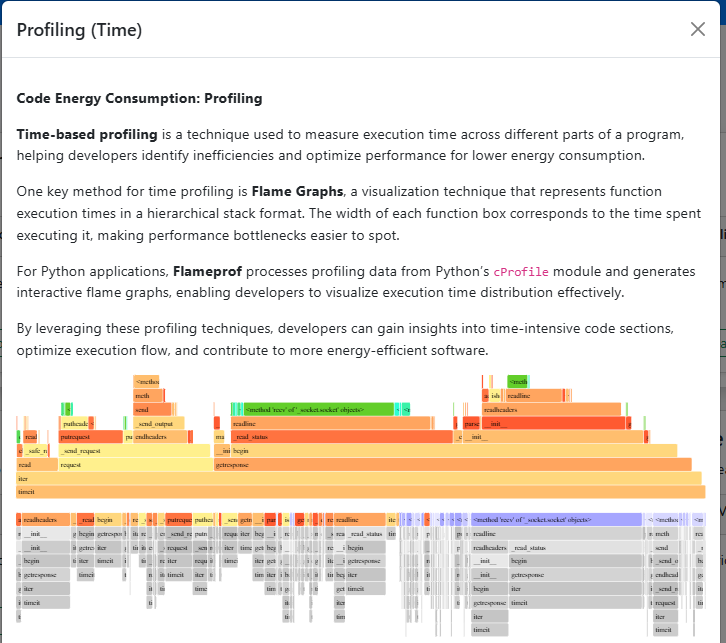}
    \end{tabular}
    \caption{Left: Packaged Tool installation example for Green Refactoring. Right: Explanations of Time-Based Profiling technique to measure energy consumption.}
    \label{fig:ui_explanations}
\end{figure*}

Some well-known approaches surfaced during the literature review but were intentionally excluded due to scope. For example, Intel RAPL is a hardware interface for energy measurement rather than a developer-facing tool. It is therefore implicitly represented through various tools across multiple categories rather than treated as a standalone tool. Similarly, tools that focus on estimating carbon emissions at the infrastructure or cloud-account level rather than measuring software energy consumption at the application or development level were considered out of scope since this study focuses on feasibility within software development workflows. 

Table~\ref{tab:sustainability_tools} contains more details on these specific categories. This allowed us to map how prior research aligns with our study's focus on installation feasibility, input feasibility, and output usefulness. These dimensions directly informed the design of workshop and survey, and the structure of our results.

Input data was characterized at a practitioner-facing level (e.g., code files, codebase access, or application execution) rather than at the level of internal measurement signals. While some energy-consumption tools do not require direct source-code access, they still operate on artifacts produced by code, such as binaries, workloads, or running processes. Abstracting inputs in this way allowed consistent comparison across tools while focusing on feasibility concerns relevant to practitioners, such as data access, compliance, and integration effort.

\subsection{Interactive Web Application}
Following the literature review, to facilitate consistent and structured evaluation of SSE tools and techniques, we developed a custom web-based application. This interface served as a central platform for both the workshop and the follow-up survey.

The web application included three primary components:
\begin{itemize}[leftmargin=*]

\item \textbf{Documentation:} Tools and techniques were divided into three groups, energy consumption, green refactoring, and workload management, and further classified as packaged, methodology-driven, or hardware-based. UI screens of these are shown in Figure~\ref{fig:ui_tools_all} and Figure~\ref{fig:ui_explanations}. Each type included an overview, input/output expectations, and installation guidance. Depending on the complexity of the tool, installation steps were provided in one of three formats: written documentation, tutorial video, or visual flow diagrams.

\item \textbf{Simulated Testing Environment:} The application featured interactive pages where participants could explore the results of applying selected tools to pre-configured code snippets. To avoid issues related to syntax errors or configuration overhead, all outputs were pre-generated and hard coded into the interface. These outputs, derived from local executions of the tools, included numerical metrics, time-series plots, and comparative visualizations, allowing participants to assess each tool's utility without needing to install or run code themselves.

\item \textbf{Feedback Interface:} Feedback forms were embedded directly within the application, covering dimensions such as tool usefulness, integration effort, installation clarity, and preferred output formats. Participants were also prompted to reflect on perceived sustainability trade-offs and organizational fit. All responses were submitted anonymously through the interface.
\end{itemize}

In addition to tools and techniques exploration, the application included a section explaining the core dimensions of software sustainability: environmental, economic, technical, individual, and social. Trade-off scenarios were presented to help participants understand how different priorities can conflict in practice, encouraging more informed and reflective evaluations of the tools.

\subsection{Workshop}
\subsubsection{\textbf{Design}}
We ran a practitioner workshop to examine the real-world feasibility of the SSE tools and techniques identified in our literature review. The workshop comprised two independent sessions: a 90-minute group discussion and an open evaluation session. Participants could choose to attend either session based on their availability and preference. The group discussion encouraged collective reflection and exchange of perspectives, while the open evaluation session allowed participants to explore the tools individually and provide feedback without group influence.

In both sessions, participants interacted with the web application described earlier, which provided some insights into sustainability and its trade-offs, and contained all tool and technique information, pre-generated output examples and embedded questionnaire forms.

During the group discussion session, teams of three to four practitioners explored the tools and responded to guided prompts focused on installation effort, input data privacy, output usefulness, and integration challenges. The open evaluation session allowed participants to engage with the tools individually and submit their reflections directly through the interface at their own pace. Prior to the main workshop, we conducted three pilot sessions with individual participants from an IT services company to check the structure, timing and clarity of the questions. These pilot participants were not included in the main study, and no changes were required.

\subsubsection{\textbf{Data Collection}}
At the start of the workshop, participants were asked to describe a real-world scenario where a sustainability-related tool or technique might be adopted. They were then prompted to reflect on potential trade-offs across five dimensions of sustainability: environmental, economic, technical, individual, and social. This activity was intended solely as a warm-up task to facilitate participants' focused thinking about SSE and provide context for the tool and technique evaluations that followed. Hence, responses for this short activity were not analyzed.

After exploring each tool and technique through the web application, participants completed structured questionnaire forms embedded within the interface. These forms included open-ended questions and slider-based ratings covering perceived benefits, installation effort, data concerns, output clarity, and integration challenges. All responses were submitted anonymously through the application and stored securely for analysis.

To conclude, participants also completed the System Usability Scale (SUS) questionnaire to assess the usability of the web application itself~\cite{brooke_sus_1996}. SUS was used solely to confirm that the evaluation platform did not introduce usability-related confounds, not as a study outcome. Discussions during the group session were audio-recorded, transcribed, and anonymized. All written feedback was submitted digitally via the web application, ensuring consistent and anonymous data collection across both sessions.

Sixteen software professionals from a large multinational financial-sector organization participated in the workshop, with roles ranging from junior developer to team lead and architect. We note that 2 of the 16 participants had prior conceptual awareness of SSE but had not applied SSE practices in their work, while the remaining participants reported no prior experience and participated primarily to learn about new tools and techniques. This organization has a hierarchical software engineering workforce, where teams follow a mix of Agile and Waterfall practices. Participants were recruited on a voluntary basis through internal communication channels and networks circulated by a designated organizational gatekeeper. We provide the detailed participant demographics\footnote{Participant demographics is stored in the \texttt{Participant Demographics} folder under the name, \texttt{workshop\_participants.xlsx} in the replication package} in the replication package~\cite{anonymous_2025_18337075}.

\subsubsection{\textbf{Analysis}} 
We employed Braun and Clarke's reflexive thematic analysis method~\cite{braun_using_2006} to analyze the qualitative data collected from the workshop. This approach was chosen for its flexibility in surfacing both anticipated and emergent themes while allowing for nuanced interpretation of practitioner feedback.

To guide the initial analysis, we developed a provisional codebook grounded in the structured prompts embedded in the web application. The structure remained unchanged during the analysis. These codes reflected key evaluation dimensions: \textit{benefits}, \textit{installation feasibility}, \textit{data access concerns}, \textit{output clarity}, \textit{adoption challenges}, \textit{organizational facilitators}, and \textit{sustainability trade-offs}.

We exported all open-text responses from the interface, along with transcriptions of the group session discussions, into a single document. The data were imported into Microsoft Excel for systematic coding. Using a deductive approach, responses were reviewed line-by-line and coded iteratively by one researcher, while the other two served as peers to simulate peer debriefing. The peers reviewed the themes by questioning interpretations and challenging assumptions. Coding was refined iteratively through discussion until all three researchers reached consensus on the final thematic structure.

The resulting thematic structure maps directly to the dimensions explored in the follow-up survey. This alignment ensured that the qualitative insights from the workshop could be further validated and elaborated through broader, quantitative data collection.

In addition to thematic coding of the qualitative data, we performed exploratory quantitative analysis on the structured slider-based ratings that participants submitted through the workshop interface. For each dimension (installation feasibility, data access, and output usefulness), we calculated descriptive statistics such as mean, median, standard deviation, and range to capture overall trends. To better illustrate variation and participant preferences, we used violin plots and ranking heatmaps (see replication package~\cite{anonymous_2025_18337075}\footnote{The replication package~\cite{anonymous_2025_18337075} includes the heatmaps in the \texttt{screenshots} folder: \texttt{workshop\_heatmap.png} and \texttt{survey\_heatmap.png}.}). Violin plots were chosen because they provide a compact way to show both the distribution and density of responses, which is particularly useful when working with limited data points, while ranking heatmaps allowed us to highlight relative preferences across dimensions in a visually accessible manner. Finally, we examined potential associations between dimensions (e.g., installation method preferences and favored output types) using Spearman rank correlation, as it is well-suited for small datasets and ordinal rating scales. Unlike other tests, Spearman does not assume linearity or normally distributed data, making it more appropriate for our exploratory analysis of slider-based ratings. Given the limited sample size and the ordinal nature of the slider-based ratings, these results are intended to highlight potential patterns and relationships rather than to serve as formal hypothesis tests. We therefore report correlations as descriptive evidence to guide future, larger-scale investigations.

To assess whether the evaluation platform itself introduced usability concerns or information-overload, participants completed the System Usability Scale (SUS) at the end of the workshop. The mean SUS score was 77.8 (SD=13.9, N=17), which falls within the good to excellent range based on standard interpretation guidelines. This suggests that participants were able to navigate the web application and engage with the tool examples without substantial usability barriers, supporting the validity of the collected feedback.

\subsection{Survey}
\subsubsection{\textbf{Design}}
Following the workshop, we launched an anonymous online survey to collect inputs from software practitioners beyond the financial sector. While the workshops offered detailed qualitative and quantitative insights, the survey aimed to compare and validate the findings at scale by focusing on key practical aspects of SSE tool adoption.

To develop the survey we drew directly on the themes that emerged from the workshop discussions and written feedback. Open ended prompts from the workshop were transformed into structured questions so that the same ideas could be examined quantitatively. For example the open question asking participants to describe the conditions that would encourage adoption was converted into a multiple choice item where respondents selected the top three factors that would increase the adoption of SSE tools in their organization such as low or no cost automation support, IDE or pipeline compatibility, strong compliance and privacy support training availability, senior leadership endorsement, quick setup metrics and dashboards for reporting and integration with legacy systems. An additional open ended text box allowed participants to share any other factors or thoughts that were not covered by the predefined choices.

The survey was distributed through practitioner-focused channels, including industry forums, academic mailing lists, and professional networks. It was open to anyone with a background in software development and designed to take approximately 10–20 minutes to complete.

Unlike the workshop, survey participants were presented with all the questions upfront and could optionally explore the web application if needed. This reversed interaction flow allowed for flexible engagement depending on each participant's interest or familiarity with the tools.

\subsubsection{\textbf{Data Collection}}
The survey design was deliberately narrow and structured around the key barriers, trade-offs, and usability factors observed during the workshop sessions. The survey comprised the following sections.
\begin{itemize}[leftmargin=*]
    \item \textbf{Background Information:} Participants reported their sector, organization size, job role, years of experience, and the types of development environments they work with (e.g., IDEs, CI/CD pipelines, cloud).
    
    \item \textbf{Installation Feasibility:} Participants rated how feasible it would be to integrate SSE tools installed via different methods (e.g., IDE plugins, direct installs, manual builds, hardware devices), using (1-10) slider-based ratings.
    
    \item \textbf{Data Sensitivity and Compliance:} Questions captured which types of data participants' organizations allow tools to access (e.g., source code, code base, runtime) and their level of concern around compliance, data exposure, and added complexity.
    
    \item \textbf{Usefulness of Outputs:} Participants rated the usefulness of output types, including raw sustainability metrics, refined suggestions, auto-refactoring hints, and dashboards.
    
    \item \textbf{Adoption Factors and Support:} Final questions captured top factors influencing tool adoption (e.g., low cost, privacy support, training, executive buy-in), along with open responses on challenges and desired support.
\end{itemize}

\subsubsection{\textbf{Screening}}
To ensure data quality, responses were excluded if participants indicated they were not in a software practitioner role or completed less than 40\% of the survey. The survey remained open for 30 days and received a total of 42 responses and after screening, 27 responses were considered valid.

\subsubsection{\textbf{Analysis}}
The survey data was analyzed using the same quantitative procedures as the workshop, since both collected slider-based ratings on comparable dimensions. This allowed us to compute descriptive statistics and generate violin plots, heatmaps$^3$, and correlation analyses consistently, making direct comparisons between workshop and survey findings possible. For open-ended responses and categorical-option questions, the data volume was relatively small, so these responses were coded directly as themes rather than integrated into a broader codebook.

\section{Results}
We organized the insights obtained from the survey and workshop results into five key topics. 

\subsection{Installation Feasibility}

Participants strongly preferred tools that could be embedded within existing IDEs or CI/CD pipelines. On a 1-10 scale, IDE plugins received the highest feasibility rating (M=9.6, SD=0.8), while manual builds (M=3.8, SD=3.1) and hardware devices (M=2.8, SD=2.8) were viewed as least practical. One participant noted, \textit{``Plugins inside currently used tools is quite useful ... if it is something that can fit in [a] pipeline which can be automated, I think that's the most useful and has [a] higher chance of being implemented.''}

Participants repeatedly highlighted the long approval procedures for installing new software in their organizations, noting that even simple executables required IT sign-off. A minority suggested that hardware additions may sometimes be easier to approve than new software, although they acknowledged the overhead in setup. This reflects a broader challenge where feasibility is shaped more by organizational processes than by technical difficulty.

\begin{figure*}
    \centering
    \includegraphics[width=0.75\textwidth]{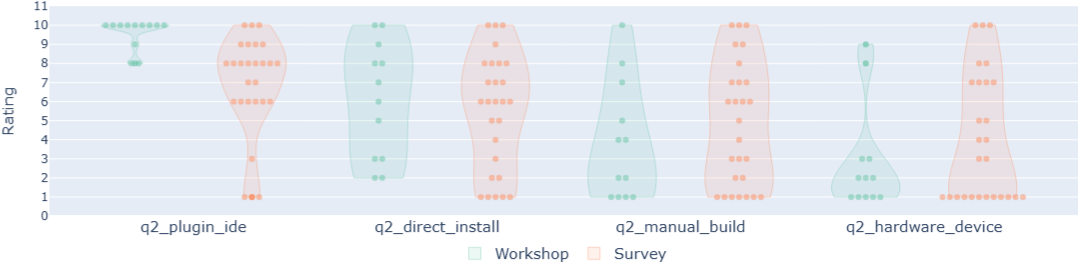}
    \caption{Violin plots showing feasibility ratings (1–10) for different installation methods. The colored regions indicate the distribution and density of responses. Each dot represents an individual participant's rating. IDE plugins received the highest feasibility ratings in both datasets, while manual builds and hardware devices were consistently rated lowest.}
    \label{fig:violin_installation}
\end{figure*}

Survey results broadly mirrored the workshops, again favoring IDE plugins as the most practical integration method and hardware devices as least feasible. Workshop and survey responses' comparison through violin chart can be seen in Figure~\ref{fig:violin_installation}. The difference lay in overall ratings: while financial-sector participants in the workshops rated plugins extremely highly (M=9.6), survey respondents from more varied domains were more cautious, averaging 6.8. Direct installs received more mixed ratings, while manual builds and hardware devices were consistently low. This suggests that, despite shared preferences in ordering, perceptions of feasibility vary with organizational context and installation policies.

\begin{figure*}
    \centering
    \includegraphics[width=0.75\textwidth]{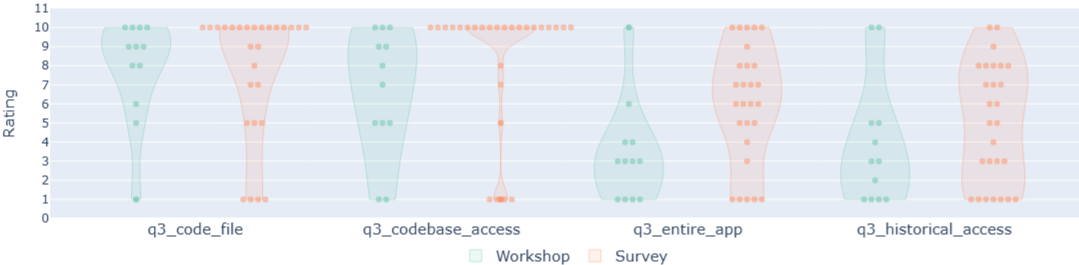}
    \caption{Violin plots showing feasibility ratings for different input data types. Workshop participants favored access to individual code files. Survey respondents rated full codebase access as most feasible and then individual code files.}
    \label{fig:violin_input}
\end{figure*}

\begin{figure*}
    \centering
    \includegraphics[width=0.75\textwidth]{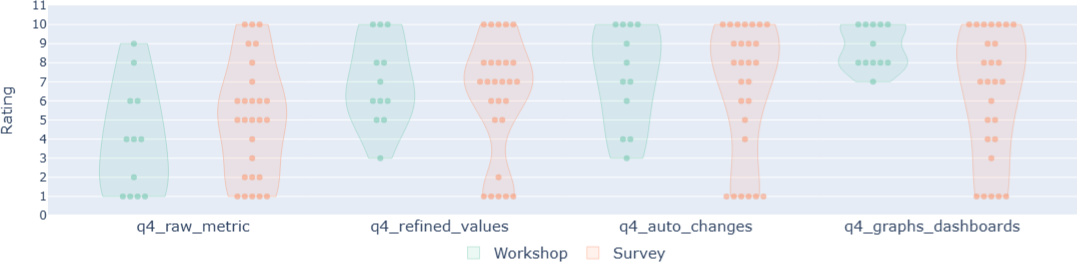}
    \caption{Violin plots showing feasibility ratings for different output formats. Dashboards and automated refactoring suggestions were perceived as most useful. Raw metrics were rated lowest, particularly by workshop participants.}
    \label{fig:violin_output}
\end{figure*}

\subsection{Input Feasibility}

Concerns about data access and compliance were paramount. Accessing individual code files was considered feasible (M=7.9, SD=2.7), while full application access (M=3.3, SD=2.6) and historical data (M=3.8, SD=3.2) were considered infeasible. As one participant explained, \textit{``If everything is running internally, then current code file analysis is totally fine. But when it comes to [a] codebase which has the data, it is not feasible, as exposing data is against compliance.''}

Practitioners stressed that all data must remain on-premises or within firewalls. Tools requiring data uploads outside the organization were deemed unacceptable, even if anonymized. Several participants also expressed hesitation about tools that require access to the entire codebase, reflecting both compliance restrictions and concerns about unintended data exposure.

Survey findings aligned with the workshops in viewing restricted project-data access as generally feasible. Figure~\ref{fig:violin_input} shows the comparison of workshop and survey responses through violin charts. The divergence arose in how the participants viewed broader data access: unlike financial-sector practitioners in the workshops, who were reluctant to grant tools access to the full codebase (M = 3.3), survey respondents rated this option much more favorably, with codebase access scoring highest overall (M = 8.3) and code file access close behind (M = 7.7). A likely explanation is that the workshop's financial-sector context is characterized by strict data security and compliance policies, whereas the survey captured perspectives from a wider range of organizations, some of which operate with more flexible data governance and mature infrastructure, making them more open to broader data integration.

\subsection{Output Feasibility}

Outputs that were visual, interpretable, and actionable were valued most. Dashboards scored highest (M=8.8, SD=1.1), followed by auto-refactoring suggestions (M=7.3, SD=2.6). Raw metrics alone were rated low (M=3.9, SD=2.8). One participant summarized this preference: \textit{``People love graphs, they love dashboards, they love green numbers. So, visualizations sounds like a very good idea.''}

However, participants also warned that graphs without context are not useful: \textit{``We can have a graph, but what does it mean? That we don't know.''} Automated code refactoring was seen as particularly attractive because it reduced developer burden: \textit{``Automated code refactoring is very helpful, because it gives alternatives of what can be done, rather than just flagging not sustainable.''} These findings suggest that interpretability and actionable feedback are as important as accuracy in shaping the perceived value of SSE tools.

Both the workshop and survey findings agreed that auto-refactor-ing and dashboards are the most useful outputs, followed by actionable suggestions like refined values. A comparison of workshop and survey responses through violin charts can be seen in Figure~\ref{fig:violin_output}. The divergence lay in perceptions of raw metrics: while workshop participants rated them low in usefulness, survey respondents gave them a moderate score of 5.03. This difference may be explained by contextual factors. In the workshop, practitioners emphasized the need for interpretable outputs that could be immediately acted upon in time-constrained contexts, whereas survey participants appeared more open to a wider range of outputs, including raw metrics, when considering potential tool adoption.

\begin{table}[h]
\centering
\caption{Summary statistics of installation feasibility, input feasibility, output usefulness (1--10 scale), and usability (SUS).}
\label{tab:results_summary_stats}
\resizebox{\columnwidth}{!}{%
\begin{tabular}{l c c c|c c c}
\hline
 & \multicolumn{3}{c}{\textbf{\textit{Workshop}}} & \multicolumn{3}{c}{\textbf{\textit{Survey}}} \\
\textbf{Category} & \textbf{Mean} & \textbf{Median} & \textbf{Mode} & \textbf{Mean} & \textbf{Median} & \textbf{Mode} \\
\hline
\multicolumn{7}{l}{\textit{Installation Feasibility}} \\
IDE plugin & 9.6 & 10.0 & 10.0 & 6.9 & 8.0 & 8.0 \\
Direct install & 6.1 & 6.5 & 2.0 & 5.5 & 6.0 & 1.0 \\
Manual build & 3.8 & 3.0 & 1.0 & 4.8 & 5.0 & 1.0 \\
Hardware device & 2.8 & 2.0 & 1.0 & 4.1 & 3.0 & 1.0 \\
\hline
\multicolumn{7}{l}{\textit{Input Feasibility}} \\
Code file & 7.9 & 9.0 & 10.0 & 7.7 & 10.0 & 10.0 \\
Codebase access & 6.7 & 7.5 & 5.0 & 8.3 & 10.0 & 10.0 \\
Entire application & 3.3 & 3.0 & 1.0 & 6.0 & 6.0 & 1.0 \\
Historical data & 3.8 & 3.0 & 1.0 & 5.0 & 5.0 & 1.0 \\
\hline
\multicolumn{7}{l}{\textit{Output Usefulness}} \\
Raw metrics & 3.9 & 4.0 & 1.0 & 5.0 & 5.0 & 1.0 \\
Refined values & 7.0 & 6.5 & 6.0 & 6.2 & 7.0 & 7.0 \\
Dashboards & 8.8 & 8.5 & 8.0 & 6.4 & 7.0 & 10.0 \\
Auto-refactoring & 7.3 & 7.5 & 10.0 & 6.6 & 8.0 & 10.0 \\
\hline
\multicolumn{4}{l}{\textit{Interface Usability}} \\
System Usability Scale (SUS) & 77.8 & -- & -- & -- & -- & -- \\
\hline
\end{tabular}
}
\end{table}

\subsection{Adoption Factors}
Beyond technical features, several organizational and cultural factors shaped adoption likelihood. Automation and minimal manual effort were critical for uptake, with developers showing reluctance toward tools that add extra overhead. Time and training constraints emerged as barriers: \textit{``We barely get time to test applications at the moment, timelines/budgets so tight.'' } 

Management buy-in was repeatedly emphasized: \textit{``Organizational change is required \ldots buy in from senior colleagues would make the application of these tools and techniques more feasible.''} Practitioners identified facilitators such as tech leads, IT services, agile coaches, and CTO-level champions who could mandate or streamline adoption. These findings highlight that SSE tools will only gain traction if aligned with organizational priorities and supported by leadership.

Survey responses reinforced workshop findings that compliance assurance, automation, and management buy-in are essential for adoption. However, they placed additional emphasis on factors not strongly voiced in workshops, such as low or no cost and the availability of structured training. Written comments highlighted data privacy and tool compatibility as recurring concerns, while suggestions for support focused on vendor-led workshops and integration demonstrations. These differences likely reflect the broader survey audience, where resource constraints and diverse organizational cultures shape perceptions of adoption feasibility.

\subsection{Cross-Dimensional Correlations}

To understand how perceptions of installation, input, and output feasibility are interrelated, we computed Spearman correlations across the three dimensions for the workshop and survey$^3$.

In both datasets, three patterns emerged. First, installation and input feasibility were positively correlated, particularly for IDE plugins and code file access~(spearman's $\rho = 0.53$ for code file access and $\rho = 0.48$ for codebase access). This suggests that participants who favored lightweight installation modes also tended to view lightweight data inputs as more feasible.

Second, input and output feasibility showed strong positive associations. Practitioners who were more open to broader inputs (e.g., codebase or historical access) also tended to rate richer outputs such as dashboards~($\rho=0.70$ and $\rho=0.73$, respectively) and automated changes~($\rho=0.74$ for code file access) as more useful.

Finally, installation and output feasibility were modestly correlated. Participants who rated plugins and direct installs highly also gave higher ratings to interpretable outputs such as dashboards~($\rho = 0.64$) and automated changes~($\rho = 0.54$). In contrast, hardware devices showed weak or even negative associations with useful outputs, aligning with their overall low feasibility ratings.

These findings show that installation, input, and output feasibility are linked: participants who preferred easy installation also favored minimal data inputs and clear outputs. We note that this correlation analysis does not demonstrate statistical significance, but rather complements the descriptive findings by showing how feasibility judgments across dimensions align with one another. 

\section{Discussion}
\noindent \textbf{RQ1: Feasibility of different types of SSE tools.}
The study highlights that tools demanding minimal technical effort are perceived as the most feasible for adoption. Approaches that can be introduced through lightweight installation processes and that fit naturally within existing development environments reduce disruption and the need for lengthy approval procedures. Similarly, tools that require only narrowly scoped inputs such as limited code artifacts rather than extensive historical or runtime data were viewed as more practical as they lower compliance (e.g., privacy and security) risks and minimize the effort needed to provide or manage data. By keeping both installation and input requirements low, these tools align with time-constrained workflows and security expectations.

Participants emphasized the value of outputs that are immediately interpretable and actionable. Visual summaries and automatically generated improvement suggestions were more useful than raw, uncontextualized measurements because they reduce cognitive burden and allow developers to act on insights without additional analysis. Together, these preferences show that technical feasibility is strongly shaped by the degree to which a tool minimizes overheads in setup and data handling while providing actionable results that demonstrate tangible value to practitioners. Importantly, the findings are not tied to specific tools but to feasibility characteristics that cut across tool implementations, including installation effort, required access, and output interpretability.

\vspace{4pt}
\noindent
\fcolorbox{lightblue}{darkblue}{%
  \parbox{\dimexpr\columnwidth-0.8em-2\fboxrule\relax}{%
    \vspace*{-\fboxsep}%

    \hspace*{-\fboxsep}%
    {\color{white}\bfseries
    \colorbox{lightblue}{%
      \parbox{\dimexpr\columnwidth-0.8em-2\fboxrule+0.6\fboxsep\relax}{%
        \centering Implications for Tool Designers and Researchers
      }}}
    \hspace*{-\fboxsep}%

    \vspace{-2pt}

\begin{itemize}[leftmargin=10pt]
  \item \textbf{Approval bottlenecks:} Integrate the tool into existing environments or platforms that have already been approved within the organization.
  \item \textbf{Strict data governance:} Support on-premises deployment and restrict data access (e.g., codebase) to file-level information or log files only.
  \item \textbf{Limited time budgets:} Provide auto-refactoring capabilities and graphical dashboards that require minimal technical effort to interpret.
\end{itemize}
\vspace{\fboxsep}
  }
}
\vspace{4pt}

\noindent \textbf{RQ2: Factors influencing perceptions of SEE tools' feasibility.}
Several of the preferences identified in this study, such as favoring IDE-integrated tools over manual builds or dashboards over raw metrics, are frequently discussed in practitioner blogs, industry talks, and other forms of gray literature. However, to the best of our knowledge, these observations have not been examined through systematic empirical studies that explicitly consider organizational constraints, compliance requirements, and real-world adoption contexts. Our contribution lies in providing empirical evidence and demonstrating how these preferences manifest, interact, and vary across regulated industrial contexts.

Adoption of SSE tools is shaped as much by organizational processes and culture as by technical characteristics~\cite{bamiduro_challenges_2025, ournani_reducing_2020, fatima_using_2025, montes_factors_2025, martinez_montes_evaluating_2025}. Participants described lengthy internal approval procedures and strict data governance as hurdles, where even low-effort tools could face delays if they required new permissions or access to sensitive information. Limited time and training further constrain adoption since teams often prioritize delivery goals over long-term sustainability objectives. In these contexts, tools that automate tasks and minimize effort are more likely to gain traction because they fit project pressures and do not demand extra oversight. Techniques that cannot be automated and require physical presence are more effective when practiced regularly~\cite{martinez_montes_evaluating_2025}. However, such techniques are hard to sustain without a shift in organizational priorities.

Cultural alignment and leadership support emerged as equally important. Developers noted that visible commitment from senior stakeholders such as team leads or technology executives legitimizes sustainability practices and encourages teams to allocate time for their use. Organizational champions and a clear link to strategic priorities were seen as essential to move sustainability efforts beyond individual interest. These findings underline that the practical uptake of SSE tools depends not only on technical feasibility but also on supportive organizational structures and culture where management buy-in, privacy assurances and structured support enable sustainability goals to be integrated into everyday software development.

\vspace{4pt}
\noindent
\fcolorbox{lightblue}{darkblue}{%
  \parbox{\dimexpr\columnwidth-0.8em-2\fboxrule\relax}{%
    \vspace*{-\fboxsep}%

    \hspace*{-\fboxsep}%
    {\color{white}\bfseries
    \colorbox{lightblue}{%
      \parbox{\dimexpr\columnwidth-0.8em-2\fboxrule+0.6\fboxsep\relax}{%
        \centering Implications for Policy Makers and Standards Bodies
      }}}
    \hspace*{-\fboxsep}%

    \vspace{-2pt}

\begin{itemize}[leftmargin=10pt]
  \item \textbf{Institutionalize organizational incentives} that encourage the integration of sustainability practices within software development processes.
  \item \textbf{Enforce compliance frameworks} to reduce approval bottlenecks and speed adoption of sustainable software.
  \item \textbf{Establish and maintain regulatory guidance} that supports the alignment of new technical solutions with organizational requirements, ensuring that sustainability innovations can be adopted without conflict with existing standards and policies.
\end{itemize}
\vspace{\fboxsep}
  }
}

\section{Limitations}

\textbf{Simulated tool interaction.}
Although participants explored SSE tools through an interactive web-app, tool usage was simulated. Inputs and outputs were pre-generated, and participants did not install or execute the tools themselves. As a result, challenges such as configuration effort, debugging, and runtime errors were not directly observed. This study therefore focuses on perceived feasibility at the interaction level. Future work can extend feasibility criteria to include deployment- and execution-related factors.

\textbf{Short-term, non-longitudinal exposure.}
The study captures first impressions and anticipated feasibility rather than sustained use over time. Longer-term effects such as fatigue, declining engagement, maintenance burden, or organizational drift could not be assessed. We intentionally scoped the study to early adoption decision-making, and future work can examine long-term usage as an additional feasibility dimension.

\textbf{Generalisability and industry context.}
This exploratory study involved modest sample sizes, with the workshop conducted in a single financial-sector organization and the survey including a broader but limited set of participants. Despite this, feasibility patterns across installation, input, and output dimensions were broadly consistent between the workshop and survey. Differences observed across contexts suggest that feasibility criteria are industry-specific rather than universally fixed. Further case studies across diverse industries are therefore needed to refine and compare feasibility criteria under different regulatory and organizational conditions.

\textbf{Tool and data representativeness.}
The study examined SSE tools at the level of shared feasibility dimensions (installation, input, and output) rather than evaluating individual tools in depth. While concrete tool examples were provided, the analysis intentionally focused on category-level characteristics. This abstraction was appropriate for identifying cross-cutting feasibility criteria, rather than assessing the effectiveness or maturity of specific tools.

\textbf{Facilitation and framing effects.}
Participants were provided with structured explanations of sustainability dimensions prior to evaluating the tools, which may have influenced how feasibility was interpreted. This framing was intentionally included to support a shared understanding of SSE concepts and to enable informed evaluation. During the workshop, researchers were available to clarify questions without steering participant opinions, supporting independent interpretation while reducing misunderstandings. The SUS scores further indicate that the information provided were easy to understand, suggesting no additional cognitive overhead.

\section{Conclusion and Future Works}
Sustainable Software Engineering (SSE) has attracted growing research interest, yet its integration into everyday industrial practice remains limited. Existing tools are not widely known or used, and many practitioners have only a narrow understanding of how such tools can fit into their workflow. This study begins to address that gap by examining how practitioners evaluate the feasibility of SSE tools in their own contexts and shows that perceptions of feasibility are highly dependent on organizational setting.

Our workshop involved 16 practitioners from a financial-sector organization, and the follow-up survey had 27 valid responses. As an exploratory study, the statistical claims are indicative rather than definitive. The financial sector's strict regulatory and data-governance requirements likely heightened concerns about tool installation, data access and integration risks. Although the survey reached a broader set of organizations, this regulation-heavy context should be kept in mind. Less regulated industries may find some SSE tools, especially those requiring broad codebase access or cloud-based services, more feasible. Case studies across diverse industries and cross-industry comparative studies will help validate and refine the feasibility criteria.

This study focuses on deriving practitioner-informed feasibility criteria rather than evaluating or ranking individual SSE tools. While concrete tool examples were used to ground discussions, no systematic comparison was conducted between tools that have achieved industrial adoption and those that remain primarily academic. Future work can build on the feasibility criteria identified in this study by conducting a comprehensive review of existing SSE tools and assessing their alignment with these criteria. Such work could compare tools that are actively used in practice with those that have seen limited adoption, examining how design decisions around installation, input requirements, and output formats influence real-world requirements.

Our literature review showed that many identified SSE tools are AI-based, typically requiring complex setup, large volumes of internal training data, and often with limited accuracy, making practical adoption challenging. Future work should develop SSE tools that reflect these technical, organizational, and cultural considerations. Action research and longitudinal studies can bring sustainability research closer to everyday software engineering practice.

\balance
\bibliographystyle{ACM-Reference-Format}
\bibliography{citations}

\end{document}